%% file: flow.tex
\documentclass[11pt]{article}

\usepackage[margin=1.5in]{geometry}
\usepackage{graphicx}
\usepackage{caption}
\usepackage{subcaption}
\usepackage{amsmath}
\usepackage{amsfonts}
\usepackage{amssymb}
\usepackage{authblk}

\input{latex-commands-new}

\usepackage{booktabs}
\usepackage{url}

\author[1]{Graham McNeill}
\author[1]{Jonathan Bright}
\author[1]{Scott A.~Hale}
\affil[1]{Oxford Internet Institute, University of Oxford}

\title{Estimating Local Commuting Patterns From Geolocated Twitter Data}
\begin{document}
\maketitle

\section*{Abstract}
The emergence of large stores of transactional data generated by increasing use of digital devices presents a huge opportunity for policymakers to improve their knowledge of the local environment and thus make more informed and better decisions. A research frontier is hence emerging which involves exploring the type of measures that can be drawn from data stores such as mobile phone logs, Internet searches and contributions to social media platforms, and the extent to which these measures are accurate reflections of the wider population. This paper contributes to this research frontier, by exploring the extent to which local commuting patterns can be estimated from data drawn from Twitter. It makes three contributions in particular. First, it shows that simple heuristics drawn from geolocated Twitter data offer a good proxy for local commuting patterns; one which outperforms the major existing method for estimating these patterns (the radiation model). Second, it investigates sources of error in the proxy measure, showing that the model performs better on short trips with higher volumes of commuters; it also looks at demographic biases but finds that, surprisingly, measurements are not significantly affected by the fact that the demographic makeup of Twitter users differs significantly from the population as a whole. Finally, it looks at potential ways of going beyond simple heuristics by incorporating temporal information into models. 

\section*{Introduction}\label{sec:intro}
Population movement is a key issue for contemporary policymakers, who need to optimise transport infrastructures and services which are under ever increasing pressure. However these policymakers often operate in an information scarce environment: existing measurement instruments such as transport surveys (which may involve stopping people as they cross the border or drive through a specific road) are time consuming and expensive as well as being a source of frustration for the local population. Hence data collection is infrequent and decisions are often based on incomplete, out-of-date estimates.

The emergence of social media as a potential window on population movement offers a huge opportunity in this regard \cite{Mayer-Schonberger2013,Lazer2009,Voigt2016}. Data from social media platforms is often available to policymakers at relatively low cost, sometimes even for free. It can also be sourced without creating disruption, as the data is generated as a byproduct of interaction with the social media platform. Furthermore, it is available in huge quantities on an ongoing basis, meaning that real time changes can be tracked. Hence the potential of social media as a secondary data source is considerable.

The availability of social media data has made it a rich area for academic research on the extent to which it can offer census like indicators in a whole variety of areas \cite{Yasseri2016, Yasseri2014}. Some initial work has started to emerge in the area of population movement in particular. For example Liu et al. \cite{Liu2014} have investigated the correlation between population and Twitter data in Australia, finding that at large scales population levels could be estimated from the prevelance of geolocated Twitter content. Noulas et al.~\cite{noulas2012tale} analysed Foursquare check-in data made available via Twitter to describe intra city movement patterns in 34 different locations in the US. Hawelka et al. \cite{Hawelka2013} meanwhile, looked at international population mobility, tracing the number of international tourists, again based on geolocated Twitter data; whilst Beiró et al. took a similar approach to domestic commuting patterns \cite{Beiro2016}. However, while these initial results are promising, much more remains to be done in terms of understanding the extent to which social media can be used systematically as an accurate indicator of population movement.

The aim of this article is to build on this existing literature by examining the extent to which local commuting patterns in the United Kingdom can be inferred from data sourced from Twitter. We make three contributions in particular. First, we show that simple heuristics applied to geolocated Twitter data offer a good proxy for local commuting patterns; one which outperforms the major existing method for estimating these patterns (the radiation model). Second, we investigate sources of error in our proxy measure, showing that the model performs better on short trips with higher volumes of commuters; we also look at demographic biases, and find that, surprisingly, our measurements are not significantly affected by the fact that the demographic makeup of Twitter users differs significantly from the population as a whole. Finally, we look at potential ways of going beyond our simple heuristics by incorporating temporal information into our models.

The rest of the article is structured in the following way. Section 1 sets out the methods and data, explaining the means of estimating commuting flow and also outlining the radiation model of commuting which we make use of as a benchmark. Section 2 presents the results of the model, comparing our Twitter estimates to the benchmark and also exploring sources of bias and error in the estimation. Section 3, finally, explores temporal based extensions to the model.

\section{Methods and Data}

In this section we will describe our basic approach to evaluating the extent to which commuter flows can be accurately estimated from Twitter; we also describe our data collection.

Twitter data is based on messages, known as ``tweets'', that people who are members of the social networking site send when making use of the service. Approximately 1\% of these tweets are ``geotagged'', by which we mean that they come with metainformation containing the location from which the tweet was sent \cite{Morstatter2013}. This geotagging often occurs when people send tweets from their mobile phone~\cite{graham2014}.

Geolocated tweets indicate, of course, where a user currently is, rather than anything about any journey they may make. However, it seems reasonable to assume that, whilst making use of the social network, many people may tweet from both their home and work locations. Hence a pattern of geotagged tweets, over a period of time, ought to contain information about patterns of commuting. Of course, there will be a certain amount of noise in the data as people will tweet from other locations: on their way to/from work, from restaurants, on holiday, etc. Furthermore, not all Twitter users will have a job nor will all jobs require regular commuting. One of the central questions in this article is to observe the extent to which commuting patterns can be inferred in spite of this noise.

Going from geolocated tweets to commuting patterns requires us to choose a heuristic for deciding which location is a ``home'' location and which location is a ``work'' location for each user, based on a pattern of geolocated tweets which may come from a variety of areas. We base this heuristic on a simple frequency count: the area that a user most frequently tweets from is assumed to be their home location; the second most frequent is assumed to be their work location (although in Section 3 we experiment with ways of improving on this simple method). All other locations are assumed to be areas visited which are unrelated to either living or working. To account for users who live and work in the same area, we use a threshold $\lambda$: if more than $\lambda\%$ of tweets are sent from the same area, we assume the user both lives and works in that area. In our results section, we experiment with different values of $\lambda$.

Having assigned users to home and work locations, we can construct a commuter flow matrix $\mT$:

\begin{eqnarray}
	\mT_{ij} = tw_{ij}
\end{eqnarray}

where $tw_{ij}$ is the number of users which have their home in location $i$ and work in location $j$.

We take as our ground truth dataset commuting data from the 2011 UK census\footnote{https://wicid.ukdataservice.ac.uk/}. The data concerns commuting within and between ``local authorities'', which are administrative regions within the UK. There are 378 local authorities in the UK.\footnote{In the census commuting data (and throughout this article),  Westminster and the City of London are treated as a single local authority, as are Cornwall and the Isles of Scilly.} We also use the population estimates from the 2011 census. The population of local authorities varies considerably, with the largest containing more than 1 million people and the smallest a little over 20,000 people.

In addition to comparing the Twitter data to our ground truth dataset, we also compare the accuracy of the Twitter based estimates to the accuracy of estimates produced by the existing standard in commuting flow estimation, which is the radiation model \cite{simini2012universal}. Having this comparison provides a stronger test than simply comparing the Twitter data with a ground truth dataset, as it enables us to examine the extent to which our method improves on existing freely available models. Whilst there are alternative models (\eg \cite{erlander1990gravity,noulas2012tale,lenormand2012universal}), the radiation model is well-suited to our purposes since it is parameter-free and only requires basic information about each area. Hence it offers a reasonable comparison to our context, where the aim is to infer commuting patterns with only a minimal amount of observational data.

The standard radiation model estimates the commuter flow matrix $\mT$ using:

\begin{eqnarray}
\mT_{ij} &=& c_i\operatorname{Prob}(work=j\ |\ home=i)\nonumber\\
&=&c_i\frac{n_in_j}{(n_i+s_{ij})(n_i + n_j + s_{ij})}\nonumber
\end{eqnarray}

where:

\begin{itemize}
	\item $\mT_{ij}$ is the $ij$-th entry of $\mT$, the number of commuters who live in area $i$ and work in area $j$
	\item $c_i$ is the total number of outward commuters who live in area $i$
	\item $n_i$ is the population in area $i$
	\item $s_{ij}$ is the population within a circle centered at
area $i$ and with a radius equal to the distance between areas $i$ and $j$ (the populations of areas $i$ and $j$ are not included)
\end{itemize}

The model assumes that the number of outward or external commuters from an area is proportional to its population. Hence, $c_i=C(n_i/N)$ where $C$ is the total number of outward commuters in the population and $N$ is the total population (given by $N=\sum_in_i$). If $C$ is unknown, the model can estimate $\operatorname{Prob}(work=j\ |\ home=i)$, but not commuter numbers. It is worth noting that the radiation model does not offer a prediction for internal commuting, i.e. the number of people who live and work in the same area, a point to which we will return below.

Yang \etal \cite{yang2014limits} have also introduced a 1-parameter variant of the radiation model, which has been shown to outperform the parameter-free version. We hence also include this model in our analysis. The parameter, $\alpha$, can either be calibrated from trip data or be estimated using:

\begin{equation}\label{eq:param-est}
	\alpha = {\left(\frac{l}{36[\text{km}]}\right)}^{1.33}
\end{equation}

where the area scale $l=\sqrt{area}$ is in the range 1-65 km. The estimated flow for the 1-parameter model is then given by:

\begin{eqnarray}
	\mT_{ij} = c_i\frac{\left[{(a_{ij} + n_j)}^\alpha - a_{ij}^\alpha\right](n_i^\alpha + 1)}{(a_{ij}^\alpha+1)\left[{(a_{ij} + n_j)}^\alpha +1\right]}
\end{eqnarray}

where $a_{ij} = n_i + s_{ij}$.

We will now give details of the data collected for the study. Our Twitter data covers a one year period from June 1, 2015 to May 31, 2016. This time period is not ideal, of course, as we are comparing patterns of Twitter data to the census, which took place in 2011. Nevertheless, it is the best data available for addressing the question. Using the sample stream of the Twitter API with `spritzer' level access, we collected all geotagged tweets from within a bounding box around the British Isles.%
\footnote{The coordinates used form a rectangle with a lower-left corner at -13.4139, 49.1621, and a top-right corner at 1.7690, 60.8547.}
This choice of a year period is significant: we expect (and indeed we found) that user engagement with the platform is bursty~\cite{barabasi2005bursty}, meaning that a large time window is required to build up a consistent pattern of tweets for one user. However, this also means that we are capturing certain types of bias in our data: for example, we may pick up occasional long distance movements, such as students moving between their homes and places of study (as found by \cite{Swier2015}), which shorter time spans might avoid. We also decided to only focus on users of the platform who had a relatively high level of engagement. In particular, we discarded users that had less than 5 tweets. We also discarded users whose first and last tweets in either their detected home or work location were less than 30 days apart (to try and eliminate, for example, people who only sent a short burst of tweets from a holiday destination). This process resulted in a large number of low-intensity users being discarded. After these steps the dataset contained around half a million users.

We logged rate limiting messages from the Twitter Streaming API and found that few tweets were omitted per day due to rate limiting. We experienced no rate limiting at all for 176 days and only slight rate limiting on other days (median 4.5 tweets lost on days with rate limiting messages). Power interruptions and network connectivity resulted in additional data loss, but there is no indication of any systematic bias from these interruptions.

We assign each geolocated tweet to a local authority. This is done in one of two ways. When exact coordinates are included with the tweet, this assignment is simple, as any pair of co-ordinates will fall within only one local authority. If co-ordinates are not included, what Twitter includes instead is a bounding box around a given place or region of geographic interest (for example, a city, a county or even a country); Twitter also includes information about the type of bounding box.%
\footnote{It is worth noting that the time period considered starts after Twitter began promoting the inclusion of ``place'' in tweets rather than the exact latitude-longitude coordinates. These places are represented as bounding boxes in the data. From earlier data, we observed an 80\% decrease in exact-geotagged tweets in April 2015, though the the overall number of geotagged tweets (\ie exact or place) remained stable.}
If this type is defined as a ``city'' we use the centroid of the bounding-box as our point for geolocation, on the assumption that the majority of cities do not cross local authority boundaries
(it is worth noting that a ``city'' in this context also refers to an area of London). In total, we were able to assign 87\% of tweets to a local authority, or 122 million tweets in total. Geolocated tweets which could not be assigned are those where the area of geolocation was too high to meaningfully assign to a local authority (for example, tweets can be geolocated to ``United Kingdom'' or ``East, England'').

Of course, we do expect this assignment to contain some error within it. Users may assign any place name to a tweet: they are not required to assign the ``correct'' name. Furthermore, some bounding boxes may cross local authority boundaries, making the centroid an unreliable means of distinguishing location. Nevertheless, we expect the process to be broadly accurate. This is something supported by an observed strong correlation ($r=0.78$) between geolocated tweets and the population of each local authority (a finding which also offers further confirmation for the results from \cite{Liu2014}).

\section{Modelling Commuting with Twitter}\label{results}

We will now move on to discuss the results of the study. We will begin by outlining the accuracy of our Twitter based commuter flow predictions, compared to the radiation model. Following \cite{yang2014limits} and \cite{lenormand2012universal}, we use the ``common part of commuters" (CPC) score based on the S{\o}rensen index \cite{sorensen1948method} to assess the accuracy of commuting flow estimates. A CPC score essentially compares the similarity of two matrices, $\mL$ and $\widetilde{\mL}$, and is given by the following equation:

\begin{equation}\label{eq:cpc}
\text{CPC}(\mL,\widetilde{\mL})=\frac{2\sum\limits_{i,j=1}^K\operatorname{min}(\mL_{ij},\widetilde{\mL}_{ij})}{\sum\limits_{i,j=1}^K{[\mL + \widetilde{\mL}]}_{ij}}
\end{equation}

where $K$ is the number of rows in the matrix (in our case the number of local authority areas). CPC scores lie in $[0,1]$ with 1 indicating perfect agreement, \ie $\mL=\widetilde{\mL}$. When computing the CPC, row $i$ of both the Twitter and radiation flow matrices are normalized to sum to $c_i = C(n_i/N)$.

Table \ref{tab:Twitter-rad-results} shows the CPC values for the Twitter-based estimates, with values of $\lambda$ varying between 0.7 and 0.95, and both variants of the radiation model (using Eq \ref{eq:param-est} to estimate the parameter). As mentioned above, the radiation model does not offer an estimate for internal commuting (i.e. the number of people who live and work in the same area), whereas the Twitter model does. To properly compare the two approaches, we hence also produced a version of the Twitter model which considers only external commuting: i.e. the diagonal entries of the flow matrix are set to zero. 

Three findings are evident from the table. First, higher values of $\lambda$ are associated with higher CPC scores: in other words, classifying users as living and working in the same local authority should be done only if almost all of their tweets are coming from that one local authority. This might indicate that, typically, users make much more use of the social media network when they are at home rather than at work: hence even a small amount of tweets in another area might indicate a pattern of commuting. Secondly, and more importantly for our purposes, the results from the Twitter proxy are high, reaching above 0.7 for higher values of $\lambda$ and above 0.8 if internal commuting is included. Hence the proxy is quite good in absolute terms, especially if we consider the simplicity of the model and the data. Finally, the Twitter model outperforms both versions of the radiation model for all values of $\lambda$ considered. This shows that Twitter data can offer a good measurement of local commuting patterns that improves on existing freely available models.

\begin{table}
\begin{center}
	\begin{tabular}{*3c}
		\toprule
		{} & \multicolumn{2}{c}{CPC Scores}\\
		\cmidrule{2-3}
		\textit{Twitter Models} & All commuting & External commuting\\
		$\lambda$ = 0.70 & 0.70 & 0.67\\
		$\lambda$ = 0.75 & 0.73 & 0.68\\
		$\lambda$ = 0.80 & 0.77 & 0.69\\
		$\lambda$ = 0.85 & 0.80 & 0.70\\
		$\lambda$ = 0.90 & 0.83 & 0.71\\
		$\lambda$ = 0.95 & 0.81 & 0.71\\[0.75em]
		\textit{Radiation Models}\\
		Standard & N/A & 0.57\\
		1-parameter & N/A & 0.62\\ 
		\bottomrule
	\end{tabular}
\end{center}
\caption{CPC scores for comparisons of the Twitter model and the radiation models to commuting data from the census}\label{tab:Twitter-rad-results}
\end{table}

In addition to this general measurement of the accuracy of Twitter based proxy, it is also worth exploring some of the sources of error in the prediction. These errors are interesting from a scientific point of view, but they are also of policy relevance. If Twitter data is biased towards certain demographic groups or certain areas, then these classes may be favoured if this type of data is used in policy decisions (for example, transport infrastructures might be unknowingly adapted more toward the needs of those with a higher education level). We explore this error in a variety of ways below: in each case we make use of the best performing model identified, which was the Twitter model with internal commuting included and with $\lambda=0.9$.

\begin{figure}
    \centering
		\includegraphics[width=0.95\textwidth]{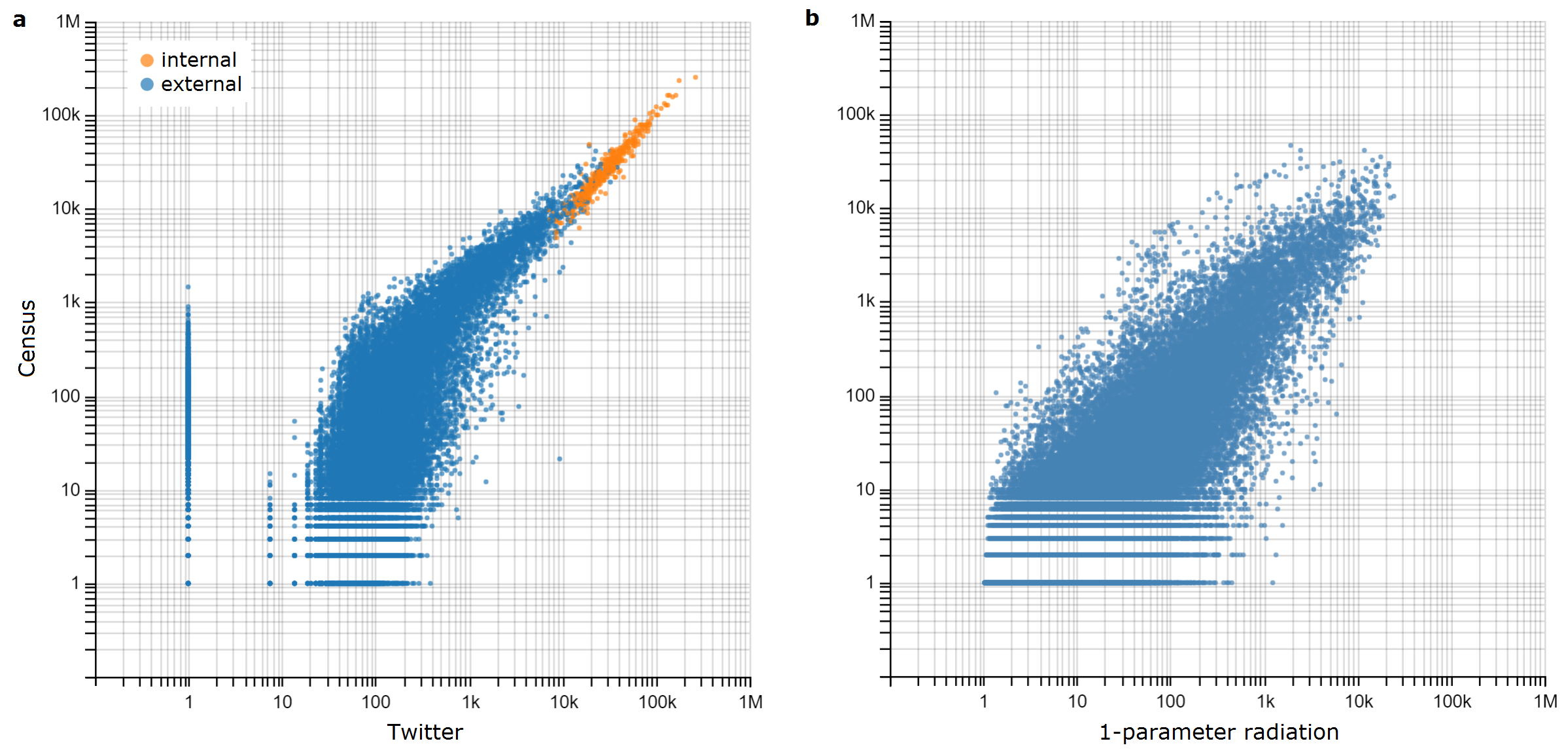}
		\caption{\textbf{Estimated commuter flow.} (a) Twitter model ($\lambda=0.9$). (b) 1-parameter radiation. Axes are log transformed, with all data points incremented by 1 before transformation. model.\label{fig:tweet-scatter}.}
\end{figure}

The first source of error we investigated was the volume of commuting between local authority pairs: we might expect local-authority pairs that share lots of commuters to be estimated better than pairs that have just a few commuters, as the signal will be stronger and hence less affected by noise in the data. Fig \ref{fig:tweet-scatter}a investigates this, by showing a plot of the estimates for the Twitter-based approach against the census data (Fig \ref{fig:tweet-scatter}b shows the estimate for the one parameter radiation model against the census data, for the purposes of comparison). We find support for this idea: both the Twitter-based approach and the radiation models become increasingly unreliable as the number of commuters decreases. Both approaches frequently predict commuting, sometimes in the hundreds, when there is none. The Twitter-based approach also frequently predicts zero commuting incorrectly. This shows that a significant portion of the CPC error concerns local authority pairs with small amounts of commuting. The greater variance in Fig \ref{fig:tweet-scatter}b (compared to that in Fig \ref{fig:tweet-scatter}a) when large numbers of commuters are involved explains the worse CPC values for the radiation models in Table \ref{tab:Twitter-rad-results}.

\begin{figure}
    \centering
		\includegraphics[width=0.65\textwidth]{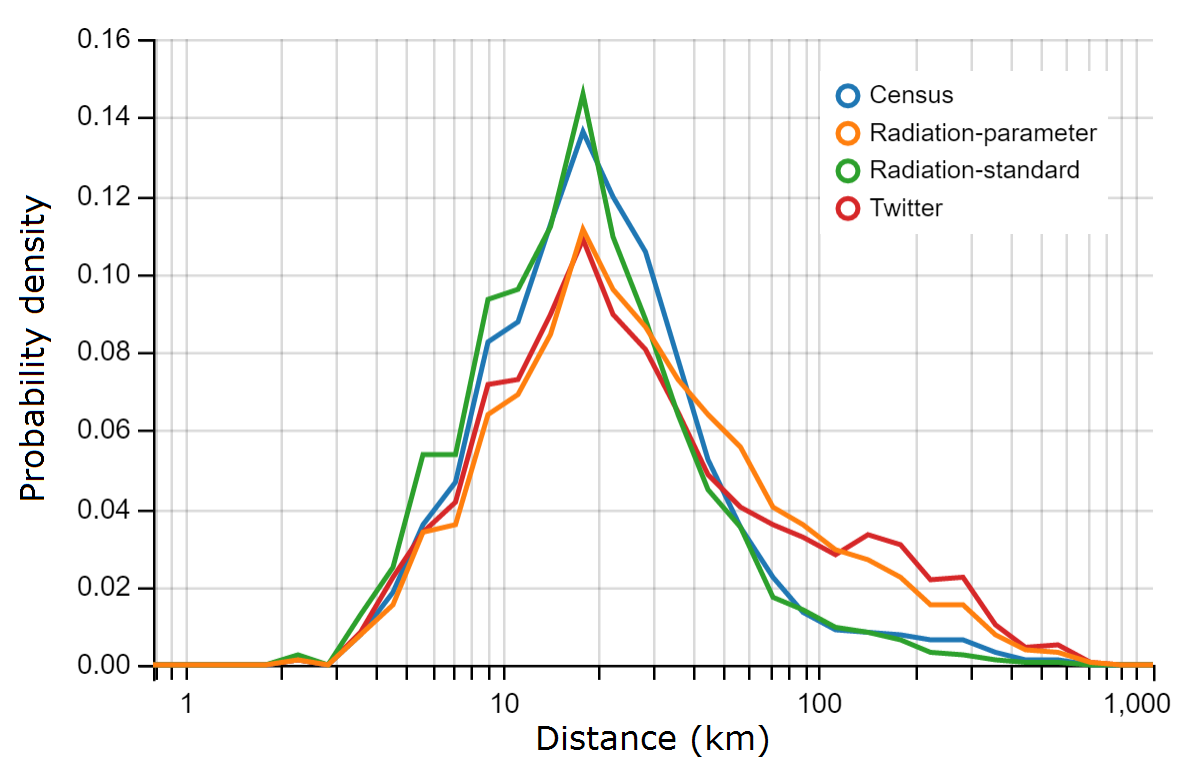}
		\caption{\textbf{Distance distribution of commuting trips.}\label{fig:distance}}
\end{figure}

A further way of investigating sources of error in the model is to look at the distribution of commuting trip distances. We might expect our poorly predicted low commuting volumes to occur disproportionately for local authority pairs that are far away from each other (who are of course likely to have fewer commuters). Fig \ref{fig:distance} shows the distribution of commuting distances (based on local authority centroids) for the census and for all of our estimates. Despite the poor CPC of the standard radiation model, it accurately estimates the distance distribution of the census. In contrast, the better performing Twitter-based approach and 1-parameter radiation model clearly overestimate the number of long commuting trips whilst underestimating the number of short ones. This would seem to connect with our observations above: Twitter is predicting small amounts of long range commuting in cases where in fact none exists. We speculate that the inaccuracy of the Twitter model in this respect stems from the fact that we are making use of one year of Twitter data, and hence may observe long term mobility patterns even whilst trying to estimate short ones: this is likely to inflate the number of long trips we estimate. This idea is supported by \cite{Swier2015}, which has shown that Twitter data can also be used to estimate long term internal migration patterns (the study focussed particularly on students leaving their home town to go to university). There is, in other words, clearly a trade off in terms of using social media data: using a longer time period allows more data to be built up, and hence potentially a stronger signal; but it also introduces new types of bias which otherwise would not be present.

We will next look at the geographic distribution of error within the models. We might expect a variety of types of geographic variation in error: for example we might observe better predictions to be available for densely populated city areas for which there is more data. In order to investigate this, we rewrite Eq \ref{eq:cpc} as:

\begin{equation}\label{eq:cpc-alt}
\text{CPC}(\mL,\widetilde{\mL})=1-\frac{\sum\limits_{i,j=1}^K\lvert\mL_{ij}-\widetilde{\mL}_{ij}\rvert}{\sum\limits_{i,j=1}^K{[\mL + \widetilde{\mL}]}_{ij}}.
\end{equation}

\begin{figure}
    \centering
		\includegraphics[width=0.65\textwidth]{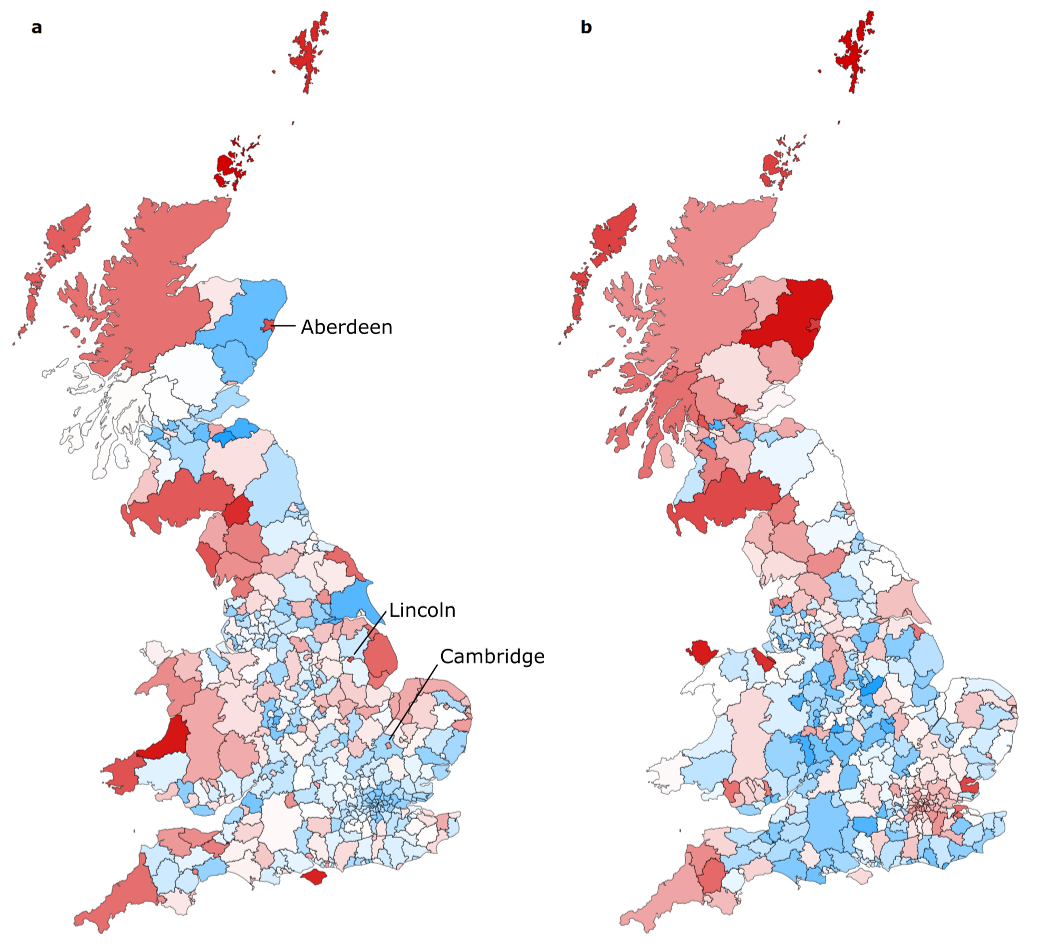}
		\caption{\textbf{Absolute error in the estimated outward commuting distributions.} The brighter the color, the greater the difference from the mean error; blue: less than mean error, red: greater than mean error. (a) Twitter ($\lambda=0.9$). (b) 1-parameter radiation model.\label{fig:maps}}
\end{figure}

The numerator in Eq \ref{eq:cpc-alt} is a sum over prediction errors; so, we can look at the error associated with a given local authority by fixing $i$ and summing over $j$ to get the outward commuting error. Since the total number of outward commuters is different for each local authority, we first normalize each row of $\mL$ and $\widetilde{\mL}$ to sum to 1 and hence, consider the outward flow probability distribution for each local authority. The results are visualized in Fig \ref{fig:maps}, which contains results both from the Twitter model and the 1-parameter radiation model for comparison.

Fig \ref{fig:maps} shows that there are some systematic patterns in the data. In the Twitter model, errors are particularly notable in medium sized cities which are surrounded by countryside (for example, the cities of Aberdeen, Lincoln and Cambridge are highlighted, all with populations between 100,000 and 300,000). This reflects the fact that the model has a hard time estimating the small amounts of commuting that go out from these cities into the surrounding countryside (by contrast, the amounts of commuting into these cities is well estimated, something shown by the fact that the areas surrounding the cities are typically blue). It is worth noting that within the large metropolitan area of London, the Twitter model performs well in absolute terms, and also much better than the radiation model, which does quite poorly in this area (something also found by \cite{Masucci2013}).  

A final area of potential error we considered concerned the impact of demographic factors. Twitter users are not representative of the general population \cite{Sloan2015,Wang2013}, and furthermore Twitter users who geotag their tweets are not even representative of Twitter users more generally~\cite{hecht2014icwsm,Malik2015}. We might expect this to distort the results; for example, Twitter commuting estimates might be better for groups which are well represented on Twitter. In order to consider the impact of demographics on commuting predictions, we make use of further census data which describes the level of commuting for a variety of different socio-demographic categories. In particular, commuting is divided up by gender, by age group, and by social class. For each of these demographic variables, we compare the performance of our predictor with the baseline performance for all types of commuting generated in Section \ref{results}.

The results of this investigation are shown in Table~\ref{tab:demog}. Differences can be observed in all the demographic categories we tested. Prediction of commuting is better for females than it is for males; it is better for the middle aged (35-64) than it is for either the young (16--34) or the old (65+). Prediction of commuting was also better for the ``lower'' social grades than the higher ones (when compared to the baseline, prediction of commuting for the AB social class was the worse of all demographics tested). However, the magnitude of the differences from the baseline in all categories is also relatively small: by and large, the difference from the baseline measure is less than 0.05 CPC. From this we infer that, even though the users of Twitter might be demographically biased, this does not hamper to a large extent our ability to infer mostly accurate commuting patterns from the data.

\begin{table}
\begin{center}
	\begin{tabular}{ l  c  c }
		\toprule		
		Commuting Type & CPC score & Difference from baseline \\
		\midrule
		All (baseline) & 0.826 & \phantom{$-$} \\
		\midrule
		\textit{Gender}  \\
		Male only & 0.786 & $-0.040$\\
		Female only & 0.842 & $+0.016$\\
		\midrule
		\textit{Age} \\
		16--24 & 0.822 & $-0.004$\\
		25--34 & 0.761 & $-0.065$\\
		35--49 & 0.818 & $-0.008$\\
		50--64 & 0.827 & $+0.001$\\
		65--74 & 0.790 & $-0.036$\\
		75\verb!+! & 0.768 & $-0.058$\\
		\midrule
		\textit{Social Class}\\
		AB & 0.721 & $-0.105$\\
		C1 & 0.805 & $-0.021$\\
		C2 & 0.813 & $-0.013$\\
		DE & 0.793 & $-0.033$\\
		\bottomrule
	\end{tabular}
\end{center}
\caption{Twitter-based estimates ($\lambda=0.9$) by gender, age and social class. Internal commuting is included.}\label{tab:demog}
\end{table}

\section{Extending the Twitter Model}\label{sec:ext}

As we have highlighted, the simple model using Twitter data offers a good proxy for local commuting patterns, but not a perfect one. In this second analytical section, therefore, we want to consider ways of extending the model. Our current model is simplistic: it makes use of a mere frequency count of the geographic locations of tweets to assign individuals to a home and work location. However, Twitter data contains much more information than just geographic locations, and including this extra information might improve our predictions. In this section, we explore one particular avenue, looking at whether temporal data might offer an improvement in accuracy. It seems reasonable to assume that tweets which are made during the working hours of the day are more likely to be from a work location (\cite{yang2014limits} has already made use of this idea when estimating mobility patterns from cell phone usage). If we include tweet timing in the model, we hence might produce a better overall estimate.

\begin{figure}
\includegraphics[width=1\textwidth]{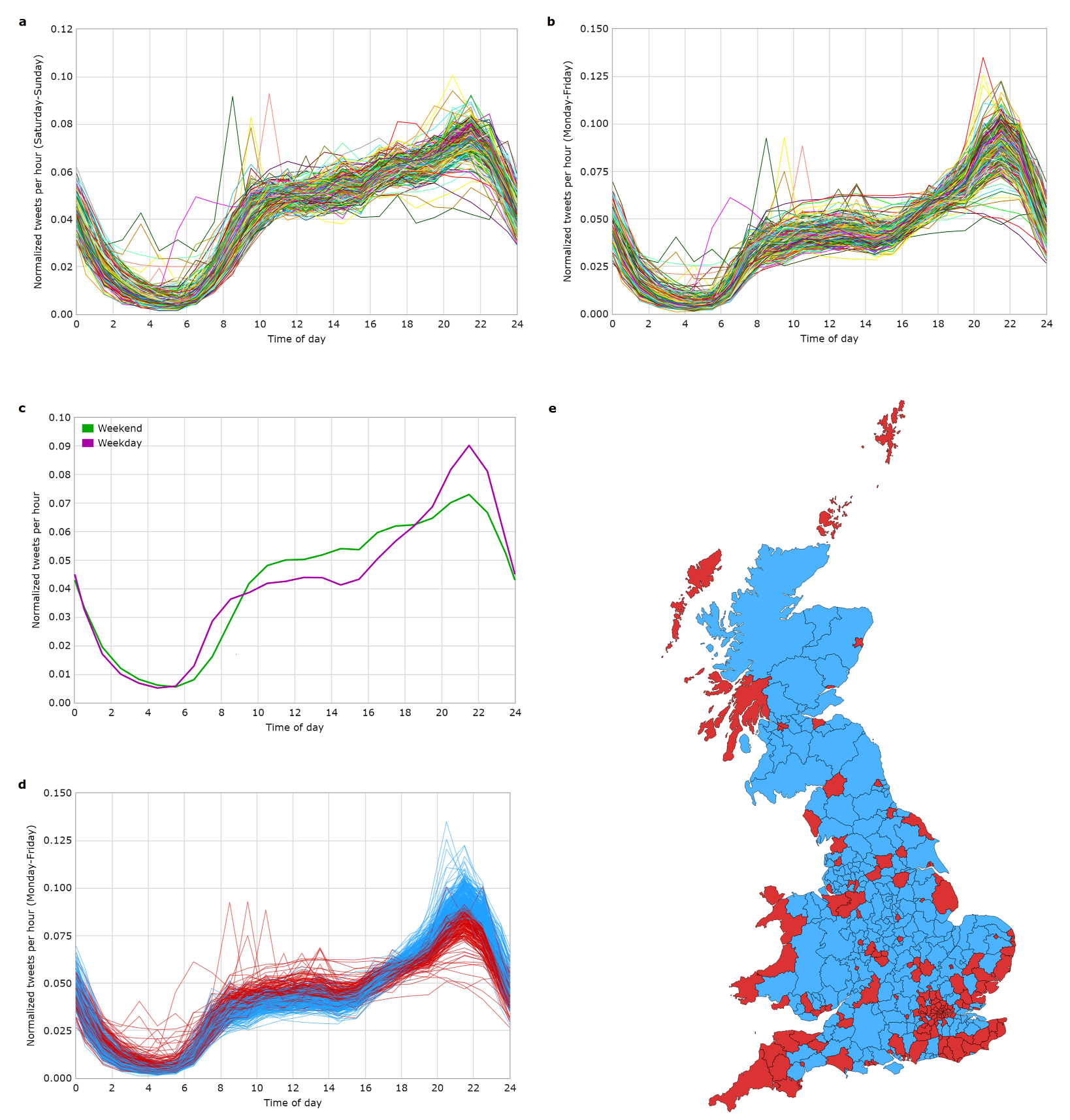}
\caption{\textbf{Temporal distribution of tweets for each local authority.} (a) Saturday and Sunday. (b) Monday-Friday. (c) Mean of weekend and weekday distributions. (d) Clustering local authorities on Monday-Friday distributions using $K$-medoids ($K=2$) with the earth-mover's distance. (e) Cluster membership of local authorities.\label{fig:temporal}}
\end{figure}

To use this assumption, we first need to define the limits of the working day. This is something we explore in Fig \ref{fig:temporal}. Figs \ref{fig:temporal}a and b shows the temporal distribution of tweets on Saturday-Sunday compared to Monday-Friday for each local authority. Fig \ref{fig:temporal}c shows the means of the weekend and weekday distributions. The Monday-Friday distributions are more consistent with typical commuting patterns: a sharp rise in tweets in the morning is followed by a consistent number of tweets during the working day and then a sharp evening peak.

Fig \ref{fig:temporal}d shows the results of clustering the distributions in Fig \ref{fig:temporal}b into two groups. One group is characterized by fewer tweets during the daytime and a higher peak in the evenings. Whilst it is inevitable that low values in one part of a distribution are balanced by higher values elsewhere, the coherence of the clusters during 10:00-15:00 and 20:00-23:00 is notable. Fig \ref{fig:temporal}e shows that throughout much of the UK, the local authorities in one of the clusters are regional hubs, which is indicated by their small size (smaller local authorities are typically those which encompass cities that have a high population density). Taken together, Fig \ref{fig:temporal} (d) and (e) are suggestive of people commuting from authorities with high outward commuting to authorities with high inward commuting.

If this is correct, these time clusters could be used to create temporal heuristics for the identification of commuting flow patterns. As a means of validating this proposition, we compute the ratio of inward to outward commuters for each authority and find the geometric mean of these ratios for each cluster.%
\footnote{People who live and work in the same local authority are not included.}
The means are 1.36 and 0.94: on average, inward commuting is 136\% of outward commuting in one cluster, but only 94\% in the other. We also computed the ratio of the hourly tweet rate between 10:00-15:00 to the hourly tweet rate between 20:00-23:00 and the ratio of inward commuting to outward commuting for each local authority. The correlation between the two sets of ratios is 0.54; the Spearman's rank correlation is 0.51. A permutation test showed that the Spearman's correlation is non-zero with p-value $<$ 0.00001. This provides good support for the idea that social media data can be used to find high and low commuting clusters; and therefore that the temporal data which is embedded within social media data might be able to improve our commuting models.

We make use of this clustering as as a way of suggesting different temporal heuristics to apply to our commuting model. We draw three inferences. First, the intersection of the clusters in Fig \ref{fig:temporal}d suggests that 09:00 and 17:00 is the typical working day. Second, the coherence of the clusters during 10:00-15:00 and 20:00-23:00 suggest that focusing on these times as a kind of restricted working day may eliminate many of the tweets sent whilst traveling to/from work and hence be optimal in terms of identifying home and work locations. Finally, the clearer signal in Fig \ref{fig:temporal}b compared to Fig \ref{fig:temporal}a suggests that we may be better using only Monday-Friday when assigning work local authorities.

By providing discrete time periods within which to assign home and work locations, the temporal approach to classification also offers a second advantage, which is that it allows us to incorporate a measure of uncertainty into the assignment of home and work locations to each user. Thus far we have made use of a simple majority rule: the location with the most tweets is assigned to the user. But theoretically there ought to be a difference in certainty according to the distribution of tweets: users who had a variety of other areas with similar amounts of tweets ought to have less certainty in them than users who had one area which had a very clear majority.

In order to incorporate the uncertainty in assigning a home and work area to each user we consider a method of soft assignment by creating a `location matrix' $\mL_u$ for each user, which is given by:

\begin{equation*}
	\mL_u = \vh_u\vw_u^T
\end{equation*}

where $\vh_u$ and $\vw_u$ are the normalized distributions of home and work tweet counts for user $u$ respectively. Hence, we can interpret the entries of $\mL_u$ as:

\begin{eqnarray}
[\mL_u]_{ij} &=& {[\vh_u]}_i{[\vw_u]}_j\nonumber\\
&=&\operatorname{Prob}(live=i)\operatorname{Prob}(work=j)\nonumber\\
&=&\operatorname{Prob}(live=i,work=j).\nonumber
\end{eqnarray}

To estimate $\operatorname{Prob}(live=i,work=j)$ for the population, we take the mean over all $U$ users:

\begin{equation*}
	\mL = \frac{1}{U}\sum_{u=1}^{U}\mL_u.
\end{equation*}

The rows of $\mL$ are then normalized as before.\footnote{Extending the filtering from Section 2, entries of $\vh_u$ and $\vw_u$ that are not associated with tweets spanning longer than a 30 day period are reset to 0 before the vectors are normalized. Users that do not have at least one non-zero entry in both $\vh_u$ and $\vw_u$ are discarded. Depending on the precise time-windows used, 75-85\% of users are discarded, leaving 287,000-496,000 `commuters'.}

Table \ref{tab:time-window} shows the CPC scores based on the temporal heuristics discussed above. Results are provided both for the Twitter model based on all commuting and the model which is based only on external commuting. For each of these two types of model, we show results using the hard assignment method and the soft assignment method which incorporates uncertainty.

The findings from this table are mixed. There is good evidence that soft assignment, which incorporates uncertainty, is better than hard assignment. There is also good evidence that focussing on Monday-Friday only gives better estimates than looking at the whole week. By contrast, there is less evidence that the time of day makes a difference, with the restricted day performing worse than the 9-5 working day in some cases. Furthermore, the overall results only offer a modest improvement on the estimations developed with simpler heuristics (see Table \ref{tab:Twitter-rad-results}), and only then in the case of external commuting. 

\begin{table}
\begin{center}
	\begin{tabular}{l *4c}	
		\toprule
		{} & \multicolumn{2}{c}{All commuting} & \multicolumn{2}{c}{External commuting}\\
		\cmidrule{2-5}
		Time Interval & Hard Assign & Soft Assign & Hard Assign & Soft Assign \\
		\midrule
		\textit{9-5 Working Day} & 0.638 & 0.699 & 0.679 & 0.729 \\
		\textit{9-5 Working Day, Mon-Fri} & 0.677 & 0.718 & 0.705 & 0.744 \\
		\textit{Restricted Day\ddag} & 0.647 & 0.697 & 0.668 & 0.721 \\
		\textit{Restricted Day, Mon-Fri} & 0.696 & 0.728 & 0.698 & 0.732 \\
		\bottomrule
	\end{tabular}
\end{center}

\caption{CPC for Twitter-based estimates using different time-based heuristics, for both hard and soft assignment techniques.\newline
\ddag The restricted day only considers tweets from 10:00-15:00 when calculating the work location, and tweets from 20:00-23:00 when calculating the home location.
}\label{tab:time-window}
\end{table}

\section{Conclusion}
In this paper we have set out to examine the extent to which Twitter data can be used to estimate local commuting flows, thus building on the nascent literature that seeks to extract reliable population indicators from social media data. We have shown that simple heuristics applied to Twitter data offer good approximations of local commuting patterns; approximations that outperform traditional estimation models such as the radiation model. We explored the sources of error in these estimations, and found that Twitter was more reliable at estimating large commuting flows over short distances, and less reliable at estimating small amounts of long range commuting. We found some evidence of geographic and demographic biases in the data, though these biases were not severe. We also explored potential extensions to the models, but in the end found that simple frequency heuristics largely outperformed more complicated models using temporal information. In conclusion, we would argue that this paper highlights again the potential of freely created and distributed social media data for understanding more about local populations.

\end{document}

%% file: latex-commands-new.tex
\newcommand{\eg}{\emph{e.g.}\ }
\newcommand{\ie}{\emph{i.e.}\ }
\newcommand{\etal}{\emph{et al.}\ }

\newcommand{\zmatrix}[1]{\ensuremath{\mathbf{#1}}}
\newcommand{\zvector}[1]{\ensuremath{\mathbf{#1}}}

\newcommand{\mL}{\zmatrix{L}}

\newcommand{\mT}{\zmatrix{T}}

\newcommand{\vh}{\zvector{h}}

\newcommand{\vw}{\zvector{w}}

%% file: flow.bbl
\begin{thebibliography}{10}

\bibitem{Mayer-Schonberger2013}
Mayer-Sch{\"{o}}nberger V, Cukier K.
\newblock {Big Data: A Revolution that Will Transform how We Live, Work, and
  Think}.
\newblock John Murray: London; 2013.

\bibitem{Lazer2009}
Lazer D, Pentland A, Adamic L, Aral S, Barabasi AL, Brewer D, et~al.
\newblock {Life in the network: the coming age of computational social
  science}.
\newblock Science (New York, NY). 2009 feb;323(5915):721--3.

\bibitem{Voigt2016}
Voigt C, Bright J.
\newblock {The Lightweight Smart City and Biases in Repurposed Big Data}.
\newblock In: The Second International Conference on Human and Social Analytics
  (HUSO 16); 2016; Forthcoming.

\bibitem{Yasseri2016}
Yasseri T, Bright J.
\newblock {Wikipedia traffic data and electoral prediction: towards
  theoretically informed models}.
\newblock EPJ Data Science. 2016;5(22).

\bibitem{Yasseri2014}
Yasseri T, Bright J.
\newblock {Can electoral popularity be predicted using socially generated big
  data?}
\newblock it - Information Technology. 2014 jan;56(5).

\bibitem{Liu2014}
Liu J, Zhao K, Khan S, Cameron M, Jurdak R.
\newblock {Multi-scale Population and Mobility Estimation with Geo-tagged
  Tweets}; 2014. Preprint. Available from: \url{http://arxiv.org/abs/1412.0327}.

\bibitem{noulas2012tale}
Noulas A, Scellato S, Lambiotte R, Pontil M, Mascolo C.
\newblock A tale of many cities: universal patterns in human urban mobility.
\newblock PloS one. 2012;7(5):e37027.

\bibitem{Hawelka2013}
Hawelka B, Sitko I, Beinat E, Sobolevsky S, Kazakopoulos P, Ratti C.
\newblock {Geo-located Twitter as proxy for global mobility patterns}.
\newblock Cartography and Geographic Information Science. 2014
  Nov;41(3):260--271.

\bibitem{Beiro2016}
Beir\'{o} MG, Panisson A, Tizzoni M, Cattuto C.
\newblock {Predicting human mobility through the assimilation of social media
  traces into mobility models}.
\newblock EPJ Data Science. 2016 Dec;5(1):17.

\bibitem{Morstatter2013}
Morstatter F, Pfeffer J, Liu H, Carley KM.
\newblock Is the sample good enough? {C}omparing data from {T}witter's
  streaming {API} with {T}witter's firehose.
\newblock ; 2013. Preprint. Available from: \url{https://arxiv.org/abs/1306.5204}.

\bibitem{graham2014}
Graham M, Hale SA, Gaffney D.
\newblock Where in the world are you? {G}eolocation and language identification
  in {Twitter}.
\newblock Professional Geographer. 2014;66(4):568--578.

\bibitem{simini2012universal}
Simini F, Gonz{\'a}lez MC, Maritan A, Barab{\'a}si AL.
\newblock A universal model for mobility and migration patterns.
\newblock Nature. 2012;484(7392):96--100.

\bibitem{erlander1990gravity}
Erlander S, Stewart NF.
\newblock The gravity model in transportation analysis: theory and extensions.  vol.~3.
\newblock Utrecht: VSP; 1990.

\bibitem{lenormand2012universal}
Lenormand M, Huet S, Gargiulo F, Deffuant G.
\newblock A universal model of commuting networks.
\newblock PloS one. 2012;7(10):e45985.

\bibitem{yang2014limits}
Yang Y, Herrera C, Eagle N, Gonz{\'a}lez MC.
\newblock Limits of predictability in commuting flows in the absence of data for calibration.
\newblock Scientific Reports. 2014;4.

\bibitem{barabasi2005bursty}
Barabasi AL.
\newblock The origin of bursts and heavy tails in human dynamics.
\newblock Nature. 2005;435(7039):207--211.

\bibitem{Swier2015}
Swier N, Komarniczky B, Clapperton B.
\newblock {Using geolocated Twitter traces to infer residence and mobility.}
\newblock Office for National Statistics GSS Methodology Series 2015;41.

\bibitem{sorensen1948method}
S{\o}rensen T.
\newblock {A method of establishing groups of equal amplitude in plant
  sociology based on similarity of species and its application to analyses of
  the vegetation on Danish commons}.
\newblock Biol Skr. 1948;5:1--34.

\bibitem{Masucci2013}
Masucci AP, Serras J, Johansson A, Batty M.
\newblock {Gravity versus radiation models: On the importance of scale and
  heterogeneity in commuting flows}.
\newblock Physical Review E. 2013 Aug;88(2):022812.

\bibitem{Sloan2015}
Sloan L, Morgan J, Savage M, Burrows R, Edwards A, Housley W, et~al.
\newblock {Who Tweets with Their Location? Understanding the Relationship
  between Demographic Characteristics and the Use of Geoservices and Geotagging
  on Twitter}.
\newblock PloS one. 2015 Nov;10(11):e0142209.

\bibitem{Wang2013}
Wang S, Lo D, Jiang L.
\newblock {An empirical study on developer interactions in StackOverflow}.
\newblock In: Proceedings of the 28th Annual ACM Symposium on Applied Computing - SAC '13; 2013.

\bibitem{hecht2014icwsm}
Hecht B, Stephens M.
\newblock A Tale of Cities: Urban Biases in Volunteered Geographic Information.
\newblock In: International AAAI Conference on Web and Social Media; 2014. 

\bibitem{Malik2015}
Malik M, Lamba H, Nakos C, Pfeffer J.
\newblock {Population Bias in Geotagged Tweets}.
\newblock In: Ninth International AAAI Conference on Web and Social Media;2015.

\end{thebibliography}
